\documentclass{aa}
\usepackage{txfonts}
\usepackage{graphicx}
\usepackage{color,enumerate}

\usepackage{natbib}
\bibpunct{(}{)}{;}{a}{}{,} 

\def\be{\begin{equation}}
\def\ee{\end{equation}}
\def\bea{\begin{eqnarray}}
\def\eea{\end{eqnarray}}
\def\d{\mbox{d}}

\def\p{\partial}

\def\pder#1#2{\frac{\partial #1}{\partial #2}}

\let\phi=\varphi
\let\rho=\varrho
\def\bh{black-hole }

\begin{document}

\title{Humpy LNRF-velocity profiles in accretion discs orbiting nearly
       extreme Kerr black holes.}
\subtitle{A possible relation to QPOs.}

\titlerunning{LNRF-velocity profiles and QPOs}

\author{Zden\v{e}k Stuchl\'{\i}k \and Petr Slan\'{y} \and Gabriel
  T\"{o}r\"{o}k} 
\institute{Institute of Physics, Faculty of Philosophy and Science, Silesian
  University in Opava, Bezru\v{c}ovo n\'{a}m. 13,
  \\
  CZ-74601 Opava, Czech Republic}

\offprints{Z. Stuchl\'{\i}k \\ \email{zdenek.stuchlik@fpf.slu.cz}}

\date{Received / Accepted}

\abstract
{Change of sign of the LNRF-velocity gradient has been found for 
  accretion discs orbiting rapidly rotating Kerr black holes with spin
  $a>0.9953$ for Keplerian discs and $a>0.99979$ for marginally stable thick
  discs. Such a ``humpy'' LNRF-velocity profiles occur just above the
  marginally stable circular geodesic of the black hole spacetimes.}
{Aschenbach (2004) has identified the maximal rate of change of the orbital
  velocity within the ``humpy'' profile with a locally defined critical
  frequency of disc oscillations, but it has been done in
  a~coordinate-dependent form that should be corrected.} 
{We define the critical ``humpy'' frequency $\nu_\mathrm{h}$ in general
  relativistic, coordinate independent form, and relate the frequency defined
  in the LNRF to the distant observers. At radius of its definition, the
  resulting ``humpy'' frequency $\nu_\mathrm{h}$ is compared to the radial
  $\nu_\mathrm{r}$ and vertical $\nu_\mathrm{v}$  epicyclic frequencies and
  the orbital frequency of the discs. 
  We focus our attention to Keplerian thin discs and perfect-fluid slender
  tori where the approximation of oscillations with epicyclic frequencies is
  acceptable.}
{In the case of Keplerian discs, we show that the epicyclic
  resonance radii $r_{3:1}$ and $r_{4:1}$ (with $\nu_{\rm v}:\nu_{\rm r}=3:1,\,
  4:1$) are located in vicinity of the ``humpy'' radius $r_{\rm h}$ 
  where efficient triggering of oscillations with frequencies $\sim\nu_{\rm h}$
  could be expected. 
  Asymptotically (for $1-a<10^{-4}$) the ratio of the epicyclic and Keplerian
  frequencies and the humpy frequency is nearly constant, i.e., almost
  independent of $a$, being for the radial epicyclic frequency
  $\nu_\mathrm{r}:\nu_\mathrm{h} \sim 3:2$.
  In the case of thick discs, the situation is more complex due to 
  dependence on distribution of the specific angular momentum $\ell$
  determining the disc properties. For $\ell=\mbox{const}$ tori and
  $1-a<10^{-6}$ the frequency ratios of the humpy frequency and the orbital
  and epicyclic frequencies are again nearly constant and independent of 
  both $a$ and $\ell$ being for the radial epicyclic frequency
  $\nu_\mathrm{r}:\nu_\mathrm{h}$ close to 4.
  In the limiting case of very slender tori ($\ell\sim\ell_{\rm
  ms}$) the epicyclic resonance radius $r_{4:1}\sim r_{\rm h}$ for all the
  relevant interval of $1-a<2\times 10^{-4}$.}
{The hypothetical ``humpy'' oscillations could be related to the QPO resonant 
  phenomena between the epicyclic oscillations in both the thin discs and
  marginally stable tori giving interesting predictions that have to be
  compared with QPO observations in nearly extreme Kerr black hole candidate
  systems. Generally, more than two observable oscillations are
  predicted.}

\maketitle 


\section{Introduction}

High frequency (kHz) twin peak quasi-periodic oscillations (QPOs) with
frequency ratios 3:2 (and sometimes 3:1) are observed in microquasars (see,
e.g.,
\citet{Kli:2000:ARASTRA:,McCli-Rem:2004:CompactX-Sources:,Rem:2005:Xray:}).  
In the Galactic Center black hole Sgr~A*,
\citet{Gen-etal:2003:NATURE:} measured a clear periodicity of 1020~sec in
variability during a flaring event. This period is in the range of
Keplerian orbital periods at a few gravitational radii from a black hole with
mass $M\sim 3.6\times 10^{6}M_{\odot}$ estimated for Sgr~A*
\citep{Ghez:2004:CarnegieObservatoriesCentennialSymposia:}. More recently
\citet{Asch-etal:2004:X-ray,Asch:2004:ASTRA:,Asch:2006:ASTROPH:} reported
three QPO periodicities at 692 sec, 1130 sec and 2178 sec that correspond to
frequency ratios $(1/692):(1/1130):(1/2178)\sim 3:2:1$. However, these
observational data are not quite convincing, see,
e.g. \citep{Abr-etal:2004:RAGtime4and5:}.
In some galactic binary black hole and neutron-star systems, the high-frequency
QPOs at $\nu_{\rm high}$ are accompanied with low-frequency QPOs at $\nu_{\rm
  low}$. The high-frequency and low-frequency QPOs are correlated and the 
ratio of the frequencies is observed to be $\nu_{\rm high}:\nu_{\rm low} \sim 
13:1$. It was first noticed by \citet{Psa-Bel-Kli:1999:ASTRJ2:} that the
correlation between high-frequencies and low-frequencies exists for black-hole
and neutron-star sources, later \citet{Mau:2002:ASTRJ2:} and
\citet{War-Wou-Pre:2003:MONNR:} extended this correlation to cataclysmic
variables and showed that it is obeyed by high-frequency quasi-coherent 
``dwarf nova oscillations'' and the low-frequency ``horizontal branch''
oscillations. At present, there is no exact model explaining the ratio $13:1$,
only a qualitative proposal exists, based on analogy with the $9$-th wave from
oceanography \citep{Abr-etal:2004:RAGtime4and5:}. In this concept, the
high-frequency QPOs are connected to transient oscillatory phenomena at random
locations in the accretion disc and are subject to the side band instability
similar to those considered in oceanography \citep{Ben-Fei:1967}. 
If a wave pulse contains initially waves of identical length and frequency
$\nu_{\rm high}$, non-linearities can cause the waves with larger amplitude to
move faster changing their wavelength. The shorter (longer) waves in front
of (behind) the pulse cause energy to concentrate at the center of the pulse
feeding thus the instability, the result of which is that every n-th wave has
a~higher amplitude creating low-frequency oscillations with frequency 
$\nu_{\rm low}\sim\nu_{\rm  high}/n$. The value of $n$ depends on details of
the hydrodynamic models and it is not fully understood in both oceanography
(where $n \sim 9$) and discography \citep[where $n \sim
12$--$14$;][]{Abr-etal:2004:RAGtime4and5:}. 

It was proposed by \citet{Klu-Abr:2001:ACTPB:} that the high frequency
  twin peak QPOs are related to the parametric or forced resonance in
  accretion discs \citep{Lan-Lif:1973:Mech:}, possibly between 
the radial and vertical epicyclic oscillations \citep{Ali-Gal:1981:GENRG2:,
  Now-Leh:1998:TheoryBlackHoleAccretionDisks:} or the orbital and one of the
epicyclic oscillations.  
These oscillations could  be related to both the thin Keplerian
discs \citep{Abr-etal:2003:PUBASJ:,Kat:2001a:PUBASJ:} or the thick, toroidal
accretion discs \citep{Rez-etal:2003:MONNR:,Klu-etal:2003::}. 
In particular, the observations of high frequency twin peak QPOs with
  the $3:2$ frequency ratio in microquasars can be explained by the 
  parametric resonance between the radial and vertical epicyclic oscillations,
  $\nu_{\rm v}:\nu_{\rm r}\sim 3:2$. This hypothesis, under the assumption of
  geodesic oscillations (i.e., for thin discs), puts strong limit on the
  mass--spin relation for the central black hole in microquasars
  \citep{Tor-Abr-Klu-Stu:2005:ASTRA:,Tor:2005:ASTRN,TOR-ETAL:2006:PRO:}.

\cite{Asch:2004:ASTRA:,Asch:2006:ASTROPH:} discovered
that two changes of sign of the radial gradient of the Keplerian orbital
velocity as measured in the locally non-rotating frame (LNRF)
\citep{Bar-Pre-Teu:1972:ASTRJ2:} occur in the equatorial plane of Kerr black
holes with $a>0.9953$. \cite{Stu-etal:2005:PHYSR4:} have found that
the gradient sign change in the LNRF-velocity profiles occurs also for
non-geodesic motion with uniform distribution of the specific angular momentum 
$\ell(r,\theta)=\mathrm{const}$ (i.e., in marginally stable thick discs)
around extremely rapid Kerr black holes with $a>0.99979$.\footnote{Note that
  the assumption of uniform distribution of the specific angular momentum can
  be relevant at least at the inner parts of the thick disc and that matter in
  the disc follows nearly geodesic circular orbits nearby the center of the
  disc and in the vicinity of its inner edge determined by the cusp of its
  critical equipotential surface \citep[see,][]{Abr-Jar-Sik:1978:ASTRA:}.}  The
global character of the phenomenon is given in terms of topology changes of
the von Zeipel surfaces (equivalent to equivelocity surfaces in the 
tori with $\ell (r,\theta)=\mathrm{const}$). Toroidal von Zeipel surfaces exist
around the circle corresponding to the minimum of the
equatorial LNRF-velocity profile, indicating possibility of development of
some instabilities in that part of the marginally stable disc with positive
gradient of the orbital velocity in LNRF
\citep{Stu-Sla-Tor:2004:RAGtime4and5:,Stu-etal:2005:PHYSR4:}. 

\begin{figure*}
\centering
\includegraphics[width=.9 \hsize]{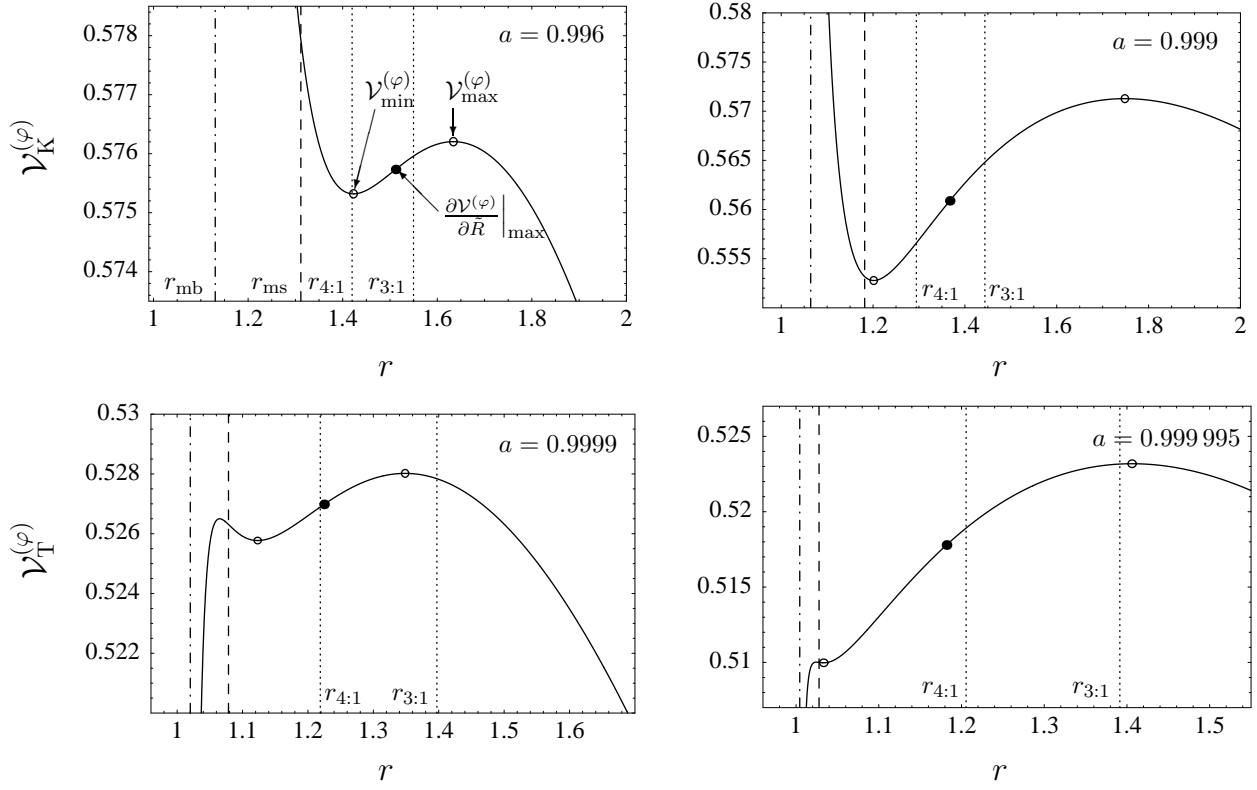}
\caption{Profiles of the equatorial orbital velocity related to LNRF
in terms of the radial Boyer-Lindquist coordinate for appropriately chosen
  values of the \bh spin $a$ in the case of Keplerian 
  discs (upper plots) and limiting marginally stable thick discs with
  $\ell=\ell_{\rm ms}=\mbox{const}$ specific angular momentum distribution
  (lower plots).
  The resonant orbits $r_{3:1}$ and $r_{4:1}$ of the epicyclic
  frequencies $\nu_{\rm v}:\nu_{\rm r}$ together with the marginally stable
  (dashed line) and marginally bound (dashed-dotted line) orbits are also 
  given.}
\label{f1}
\end{figure*}

Therefore, we consider the positive radial gradient of orbital LNRF-velocity
around black holes with $a>0.9953$, see Fig.~\ref{f1}, to be a physically
interesting phenomenon, even if a direct mechanism relating this phenomenon to
triggering the oscillations, and subsequent linking of the oscillations to the
excitation of radial (and vertical) epicyclic oscillations, is unknown. We
present a~basic study of
the ``humpy'' oscillatory frequency and its relation to the epicyclic and
Keplerian (orbital) frequencies. It should be stressed that recently at least
two QPOs sources are observed, in which the rotational parameter (spin) of the
central black hole is estimated nearly extreme, i.e., $a>0.99$. Such black
holes are probably observed in Sgr~A* \citep{Asch:2006:ASTROPH:} and in
GRS~1915$+$105 \citep{McCli-etal:2006:ASTROPH:}. We plan to make a detailed
analysis of the observed frequencies and their possible relation to the
LNRF-velocity hump induced frequency and related epicyclic frequencies in the
future work. 

\citet{Asch:2004:ASTRA:,Asch:2006:ASTROPH:}, considering phenomena observed in
Sgr A*,
has shown that in the field of the Kerr black hole with $a \simeq
0.99616$, the orbit where the critical frequency subjected to the hump
  of the LNRF-velocity profile in such a~way that the positive rate of change
  of the LNRF-velocity is maximal, $(\nu^A_\mathrm{crit}=\p {\cal
    V}^{(\phi)}/\p r)_{\rm max}$, is located nearby $r=r_{3:1}$, where
the vertical and radial epicyclic frequencies are in the ratio of $\nu_{\rm
  v}:\nu_{\rm r}=3:1$ and, moreover, the critical frequency
$\nu_{\mathrm{crit}}^A$ is nearly equal to the radial epicyclic frequency 
there. Undoubtly, this is an interesting result. However, the critical
frequency introduced by Aschenbach is related to the rate of change of the
locally measured orbital velocity in terms of the special Boyer-Lindquist
radial coordinate, so the coincidence $\nu_\mathrm{crit}^A \simeq
\nu_\mathrm{r}$ obtained in this case is rather unrealistic. In this
paper 
we give the critical frequency $\nu_{\rm crit}^{\tilde{R}}$, related to 
the maximal positive radial gradient of the LNRF-velocity in the ``humpy''
velocity profile, in the general relativistic, coordinate-independent 
form. Further, since the critical frequency $\nu_{\rm crit}^{\tilde{R}}$ is
defined locally, being connected to the LNRF, it has to be transformed into
the form related to distant stationary observers, giving observationally
relevant frequency $\nu_\mathrm{h}=\nu_{\infty}^{\tilde{R}}$.

In Section \ref{section2}, we briefly summarize properties of the Aschenbach
effect for Keplerian thin discs, and $\ell = \mathrm{const}$ thick discs. In
Section \ref{section3}, the critical frequency, connected to the LNRF-velocity
positive gradient in the humpy profiles, is given in the physically relevant,
coordinate independent form for the both Keplerian and $\ell = \mathrm{const}$
discs. At the radius of its definition, the critical frequency is compared to
the radial and vertical epicyclic frequency and the orbital frequency. In
Section \ref{section4}, the results are discussed and concluding remarks are
presented. 

\section{LNRF-velocity profiles of discs orbiting the Kerr black holes} 
\label{section2} 

In the Kerr spacetimes with the rotational parameter assumed to be $a>0$, the
relevant metric coefficients in the standard Boyer-Lindquist coordinates read:
\bea                                                        \label{e8}
     g_{tt} &=& -\frac{\Delta - a^2 \sin^2 \theta}{\Sigma}, \quad
     g_{t\phi} = -\frac{2ar\sin^2 \theta}{\Sigma},\\
     g_{\phi\phi} &=& \frac{A\sin^2 \theta}{\Sigma}, \quad
     g_{rr} = \frac{\Sigma}{\Delta}, \quad
     g_{\theta\theta} = \Sigma,
\eea
where
\bea                                               \label{e11}
     \Delta &=& r^2-2r+a^2, \quad
     \Sigma = r^2+a^2 \cos^2 \theta, \quad\\
     A &=& (r^2+a^2)^2-\Delta a^2 \sin^2 \theta.
\eea
The geometrical units, $c=G=1$, together with putting the mass of the black
hole equal to one, $M=1$, are used in order to obtain completely dimensionless
formulae hereafter.

The locally non-rotating frames (LNRF) are given by the tetrad of 1-forms
\citep{Bar-Pre-Teu:1972:ASTRJ2:}
\bea                                                          \label{e14}
     \mathbf{e}^{(t)} &=& \left(\frac{\Sigma\Delta}{A}\right)^{1/2}
     \mathbf{d}t, \quad
     \mathbf{e}^{(\phi)} = \left(\frac{A}{\Sigma}\right)^{1/2}\sin\theta\,
     (\mathbf{d}\phi-\omega\,\mathbf{d}t), \\
     \mathbf{e}^{(r)} &=& \left(\frac{\Sigma}{\Delta}\right)^{1/2}
     \mathbf{d}r, \quad
     \mathbf{e}^{(\theta)} = \Sigma^{1/2}\mathbf{d}\theta,
\eea
where
\be
     \omega = -\frac{g_{t\phi}}{g_{\phi\phi}} = \frac{2ar}{A} \label{e18}
\ee
is the angular velocity of the LNRF relative to distant observers. For matter
with a 4-velocity $U^{\mu}$ 
and angular velocity profile $\Omega(r,\,\theta)$ orbiting the Kerr black
hole, the azimuthal component of its 3-velocity in the LNRF reads 
\be                                                          \label{e19}
     {\cal V}^{(\phi)}=\frac{U^{\mu}
       \mathrm{e}^{(\phi)}_{\mu}}{U^{\nu} \mathrm{e}^{(t)}_{\nu}} =
     \frac{A\sin\theta}{\Sigma\sqrt{\Delta}} (\Omega-\omega).
\ee

\subsection{Keplerian thin discs}

In thin discs matter follows nearly circular equatorial geodetical
orbits characterized by the Keplerian distributions of the angular velocity
and the specific angular momentum (in the equatorial plane, $\theta=\pi/2$)
\bea                                                          \label{e21.1}
     \Omega&=&\Omega_{\rm K}(r;\,a)\equiv\frac{1}{(r^{3/2}+a)}, \\
     \ell&=&\ell_{\rm K}(r;\,a) \equiv
     \frac{r^2-2ar^{1/2}+a^2}{r^{3/2}-2r^{1/2}+a}\ensuremath{\mathrm{.}}
\eea
The azimuthal component of the Keplerian 3-velocity in the LNRF reads
\be                                                          \label{e22}
     {\cal V}^{(\phi)}_{\rm K}(r;\,a)=\frac{(r^2+a^2)^2 - a^2\Delta -
       2ar(r^{3/2}+a)}{r^2(r^{3/2}+a)\sqrt{\Delta}}
\ee
and formally diverges for $r\to r_{+}=1+\sqrt{1-a^2}$, where the \bh event
horizon is located. Its radial gradient is given by
\bea
    \pder{{\cal V}^{\rm(\phi)}_\mathrm{K}}{r} &=& 
    -\frac{r^5+a^4(3r+2)-2a^3
      r^{1/2}(3r+1)}{2\Delta^{3/2}\sqrt{r}(r^{3/2}+a)^2}    \nonumber \\
    &-& \frac{2a^2 r^2(2r-5)-2ar^{5/2}(5r-9)}{2\Delta^{3/2}\sqrt{r}
        (r^{3/2}+a)^2}.                                                
\eea
As shown by \citet{Asch:2004:ASTRA:,Asch:2006:ASTROPH:}, the velocity profile
has two changes of the gradient sign (where $\p {\cal V}_{\rm K}^{(\phi)}/ \p
r = 0$) in the field of rapidly rotating Kerr black holes with $a>a_{\rm
  c(K)}\doteq 0.9953$ (see Fig. \ref{f1}). 

\subsection{Marginally stable tori}
\label{s2.2}
Perfect-fluid stationary and axisymmetric toroidal discs
are characterized by the 4-velocity field $U^{\mu} =
(U^{t},\,0,\,0,\,U^{\phi})$
with $U^{t}=U^{t}(r,\,\theta),\ U^{\phi}=U^{\phi}(r,\,\theta)$, and by 
distribution of the specific angular momentum $\ell=-U_{\phi}/U_t$.
The angular velocity of orbiting matter, $\Omega=U^{\phi}/U^t$, is then
related to $\ell$ by the formula
\be                                               \label{e4}
     \Omega=-\frac{\ell g_{tt}+g_{t\phi}}{\ell g_{t\phi}+g_{\phi\phi}}.
\ee

The marginally stable tori are characterized by uniform distribution of the
specific angular momentum
\be
\ell=\ell (\mathrm{r},\,\theta)=\mathrm{const},
\ee
and are fully determined by the spacetime structure through equipotential
surfaces of the potential $W=W(r,\,\theta)$ defined by the relations
\citep{Abr-Jar-Sik:1978:ASTRA:}
\be                                               \label{e6}
     W-W_{\rm in}=\ln\frac{U_{t}}{(U_{t})_{\rm in}}, \quad
     (U_t)^2=\frac{g_{t\phi}^2 - g_{tt}g_{\phi\phi}}{g_{tt}\ell^2 +
  2g_{t\phi}\ell + g_{\phi\phi}};
\ee
the subscript ``in'' refers to the inner edge of the disc.

The LNRF orbital velocity of the torus is given by 
\be                                                          \label{e20}
     {\cal V}^{(\phi)}_{\rm T}=\frac{A(\Delta - a^2 \sin^2 \theta) + 4a^2
      r^2 \sin^2 \theta}{\Sigma\sqrt{\Delta} (A-2a\ell r)\sin\theta}\ell.
\ee

For marginally stable tori
it is enough to consider the motion in the equatorial plane,
$\theta=\pi/2$. Formally, this velocity vanishes for $r\to\infty$ and
$r\to r_{+}$, i.e., there must be
a change of its radial gradient for any values of the parameters $a$
and $\ell$, contrary to the case of Keplerian discs. The radial gradient of
the equatorial LNRF velocity of $\ell=\mathrm{const}$ tori reads 
\bea                                              \label{e23}
    \pder{{\cal V}^{(\phi)}_\mathrm{T}}{r} & = &
    \left\{\frac{[\Delta+(r-1)r][r(r^2+a^2)-2a(\ell-a)]}{[r(r^2+a^2)-2a(\ell-a)]^2\sqrt{\Delta}}\right.  \nonumber
\\
    & &\left. - \frac{r(3r^2+a^2)\Delta}{[r(r^2+a^2)-2a(\ell-a)]^2
        \sqrt{\Delta}}\right\}\ell, 
\eea
so it changes its orientation at radii determined for a given $\ell$
by the condition
\be                                               \label{e24}
     \ell=\ell_{\rm ex}(r;\,a) \equiv a +
     \frac{r^2[(r^2+a^2)(r-1)-2r\Delta]}{2a[\Delta+r(r-1)]}.
\ee

\begin{figure}
\centering
\includegraphics[width=1 \hsize]{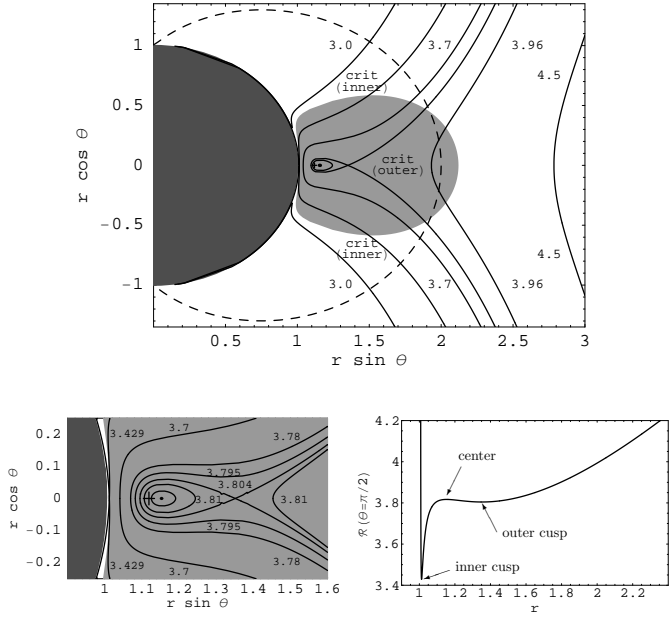}
\caption{Von Zeipel surfaces (meridional sections). For $a>a_{\rm c(T)}$
  and $\ell$ appropriately chosen, two surfaces with a cusp, or one surface
  with both the cusps, together with closed (toroidal) surfaces, exist, being
  located always inside the ergosphere (dashed 
  surface) of a given spacetime. Both the outer cusp and the central ring of
  closed surfaces are located inside the toroidal equilibrium configurations
  corresponding to marginally stable thick discs 
  (light-gray region; its shape is determined by the critical self-crossing
  {\em equipotential surface} of the potential $W(r,\,\theta)$. The cross
  ($+$) denotes the
  center of the torus. Dark region corresponds to the black hole. Figures
  illustrating all possible configurations of the von Zeipel surfaces are
  presented in \citet{Stu-etal:2005:PHYSR4:}. Here we present
  the figure plotted for the parameters $a=0.99998,\ \ell=2.0065$. Critical
  value of the von Zeipel radius corresponding to the inner and the outer
  self-crossing surface is ${\cal R}_{\rm c(in)}\doteq 3.429$ and ${\cal
  R}_{\rm c(out)}\doteq 3.804$, respectively, the central ring of toroidal
  surfaces corresponds to the value ${\cal R}_{\rm center}\doteq
  3.817$. Interesting region containing both the cusps and the toroidal
  surfaces is plotted in detail at the left lower figure. Right lower figure
  shows the behaviour of the von Zeipel radius in the equatorial plane.}
\label{f2}
\end{figure}

Of course, for both thick tori and Keplerian discs we must consider the limit
on the disc extension given by the innermost stable orbit. For Keplerian discs
this is the marginally stable geodetical orbit, $r_{\rm in}\approx r_{\rm
  ms}$, while for thick tori this
is an unstable circular geodesic kept stable by pressure gradients and
located between the marginally bound and the marginally stable geodetical
orbits, $r_{\rm mb}\lesssim r_{\rm in}\lesssim r_{\rm ms}$,
with the radius being determined by the specific angular momentum
$\ell=\mathrm{const}\in (l_{\rm ms},\,l_{\rm mb})$ through the equation $\ell=
\ell_{\rm K}(r;\,a)$; $\ell_{\rm ms}$ ($\ell_{\rm mb}$) denotes specific
angular momentum of the circular marginally stable (marginally bound) geodesic.

\begin{figure}
\centering
\includegraphics[width=.95 \hsize]{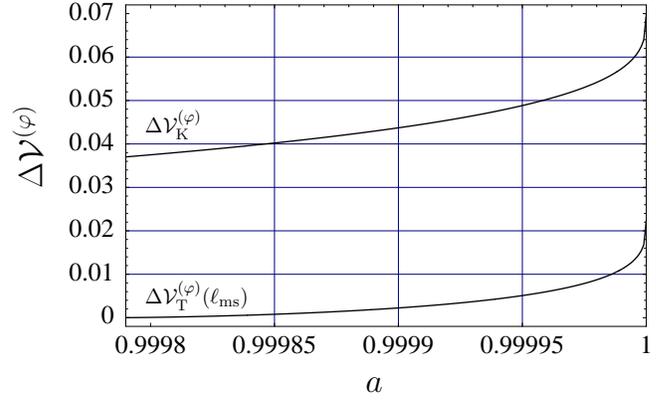}
\caption{Velocity difference $\Delta {\cal V}^{(\phi)}={\cal
    V}^{(\phi)}_{\rm max}-{\cal V}^{(\phi)}_{\rm min}$ as a function of the
    rotational parameter $a$ of the black hole for both the thin (Keplerian)
    disc and the marginally stable (non-Keplerian) disc with $\ell=\ell_{\rm
    ms}$.}
\label{f3}
\end{figure}

Detailed discussion of \citet{Stu-etal:2005:PHYSR4:} shows that
two physically relevant changes of sign of $\p {\cal
  V}^{(\phi)}_{\mathrm{T}}/\p r$ in the tori occur for Kerr black holes with
the rotational parameter $a>a_{\rm c(T)}\doteq 
0.99979$ (see Fig. \ref{f1}). The interval of relevant values of the
specific angular momentum 
$\ell\in (\ell_{\rm ms}(a),\,\ell_{\rm ex(max)}(a))$, 
where $\ell_{\rm ex(max)}(a)$ corresponds to the local maximum of the
  function (\ref{e24}), grows with $a$ growing up
to the critical value of $a_{\rm c(mb)}\doteq 0.99998$. For $a>a_{\rm c(mb)}$,
the interval of relevant values of $\ell\in (\ell_{\rm ms}(a),\ell_{\rm
  mb}(a))$ is narrowing with the rotational parameter growing up to $a=1$,
which corresponds to a singular case where $\ell_{\rm ms}(a=1)=\ell_{\rm
  mb}(a=1)=2$. Notice that the situation becomes to be singular only in terms
of the specific angular momentum; it is shown
\citep[see][]{Bar-Pre-Teu:1972:ASTRJ2:} that for $a=1$ both the total energy
$E$ and the axial angular momentum $L$ differ at $r_{\rm ms}$ and $r_{\rm mb}$,
respectively, but their combination, $\ell\equiv L/E$, giving the specific
angular momentum, coincides at these radii. 

It should be stressed that in the Kerr spacetimes with $a>a_{\rm c(T)}$,
the ``humpy'' profile of ${\cal V}^{(\phi)}_\mathrm{T}(r;\,a)$ occurs 
closely above the center of relevant toroidal discs, at radii
corresponding to stable circular geodesics of the spacetime, where the radial
and vertical epicyclic frequencies are also well defined.

A physically reasonable way of defining a global quantity characterizing
rotating fluid configurations in 
terms of the LNRF orbital velocity is to introduce, so-called, von Zeipel
radius defined by the relation
\be                                                      \label{e36}
     {\cal R}\equiv\frac{\ell}{{\cal V}^{(\phi)}_{\rm LNRF}} =
     (1-\omega\ell)\tilde{\rho},
\ee
which generalizes in another way as compared with
\citep{Abr-Nur-Wex:1995:CLAQG:} the Schwarzschildian definition of the
gyration radius $\tilde\rho$ \citep{Abr-Mil-Stu:1993:PHYSR4:}.
Note that, except for the Schwarzschild case $a=0$, the von Zeipel
  surfaces, defined as the surfaces of ${\cal R}(r,\,\theta;\,a,\,\ell) =
  \mbox{const}$, \emph{do not coincide} with those introduced by
  \citet{Koz-etal:1978:ASTRA:} as the surfaces of constant $\ell/\Omega$. 
\footnote{For more details see \citet{Stu-etal:2005:PHYSR4:}.}

In the case of marginally stable tori the von Zeipel surfaces ${\cal
    R}=\mbox{const}$ 
coincide with the equivelocity surfaces ${\cal
    V}^{(\phi)}(r,\,\theta;\,a,\,\ell)= {\cal V}^{(\phi)}_{\mathrm{T}} =
    \mbox{const}$.
Topology of the von Zeipel surfaces can be directly determined by the
behaviour of the von Zeipel radius in the equatorial plane 
\be                                                      \label{e39}
     {\cal R}(r,\,\theta=\pi/2;\,a,\,\ell) =
     \frac{r(r^2+a^2)-2a(\ell-a)}{r\sqrt{\Delta}}.
\ee
The local minima of the function (\ref{e39}) determine loci of the cusps of
the von Zeipel surfaces, while its local maximum (if it exists) determines a
circle around which closed toroidally shaped von Zeipel surfaces are
concentrated (see Fig.~\ref{f2}). Notice that the minima (maximum) of
${\cal R}(r,\,\theta=\pi/2;\,a,\,\ell)$ correspond(s) to the maxima (minimum)
of ${\cal V}^{(\phi)}_{\rm T}(r,\,\theta=\pi/2;\,a,\,\ell)$, therefore, the
inner cusp is always physically irrelevant being located outside of the
toroidal configuration of perfect fluid. Behaviour of the von Zeipel surfaces
nearby the center and the inner edge of the thick discs orbiting Kerr black
holes with $a>a_{\rm c(T)}\doteq 0.99979$, i.e., the existence of the von
Zeipel surface with an outer cusp or the surfaces with toroidal topology,
suggests possible generation of instabilities in both the
vertical and radial direction.

\begin{figure}
\centering
\includegraphics[width=.95 \hsize]{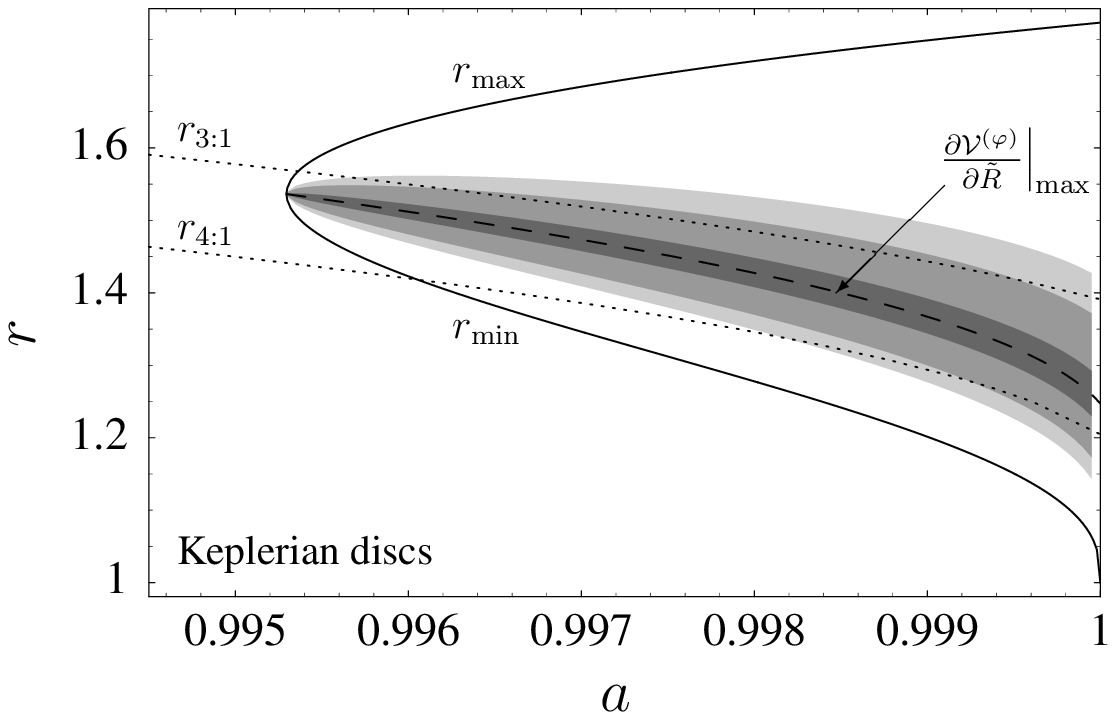}
\par\small (a)\enspace
\vskip2ex
\includegraphics[width=.95 \hsize]{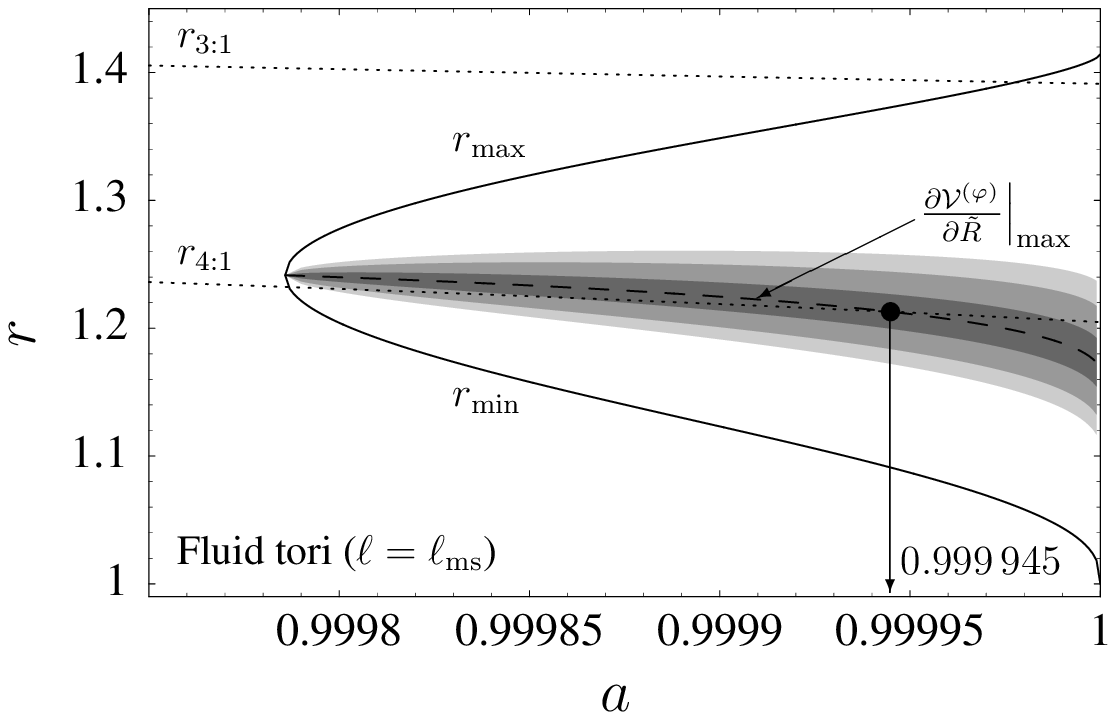}
\par\small (b)\enspace
\caption{Positions of local extrema of ${\cal V}^{(\phi)}$ (in B-L
    coordinates) for Keplerian discs (a) and marginally stable discs with
    $\ell=\ell_{\rm ms}$ (b) together with the locations of resonant orbits
    $r_{3:1}$ and $r_{4:1}$ (where the resonance between the
    vertical and radial epicyclic oscillations takes place) in dependence on
    the rotational parameter $a$ of the black hole. Dashed curve corresponds
    to the maximum positive values of the LNRF orbital velocity gradient in
    terms of the proper radial distance where the critical frequency
    $\nu^{\tilde{R}}_{\rm crit}$ is defined, boundaries of shaded regions
    correspond to orbits where the velocity gradient giving the characteristic
    frequency, $\p{\cal V}^{(\phi)}/\p\tilde{R}$, 
    reaches (a) $99\%$, $90\%$, $80\%$ and (b) $99\%$, $95\%$, $90\%$ of its
    maximum.}
 \label{f4}
\end{figure}

\subsection{Velocity profiles with a hump}
Behavior of ${\cal V}^{(\phi)}_{\mathrm{T}}(r;\,a,\,\ell)$ and ${\cal
  V}^{(\phi)}_{\rm K}(r;a)$ is illustrated in Fig.~\ref{f1}. With $a$
  growing in the region of $a\in (a_{\rm c(T)},1)$ ($a\in (a_{\rm c(K)},1)$),
  the difference $\Delta {\cal V}^{(\phi)}_\mathrm{T}\equiv {\cal 
  V}^{(\phi)}_{\rm T(max)}-{\cal V}^{(\phi)}_{\rm T(min)}$ ($\Delta {\cal
  V}^{(\phi)}_{\rm K}\equiv {\cal V}^{(\phi)}_{\rm K(max)}-{\cal
  V}^{(\phi)}_{\rm K(min)}$) grows (Fig.~\ref{f3}) as well as the 
difference of radii, $\Delta r_\mathrm{T} \equiv r_{\rm T(max)}-r_{\rm
  T(min)}$ ($\Delta r_{\rm K}\equiv r_{\rm K(max)} - r_{\rm K(min)}$), where
  the local extrema of ${\cal V}^{(\phi)}_{\rm T}$ (${\cal V}^{(\phi)}_{\rm
  K}$) occur, see Fig.~\ref{f4}.  

\begin{figure*}
\begin{minipage}{.49 \hsize}
\centering
\includegraphics[width=.95 \hsize]{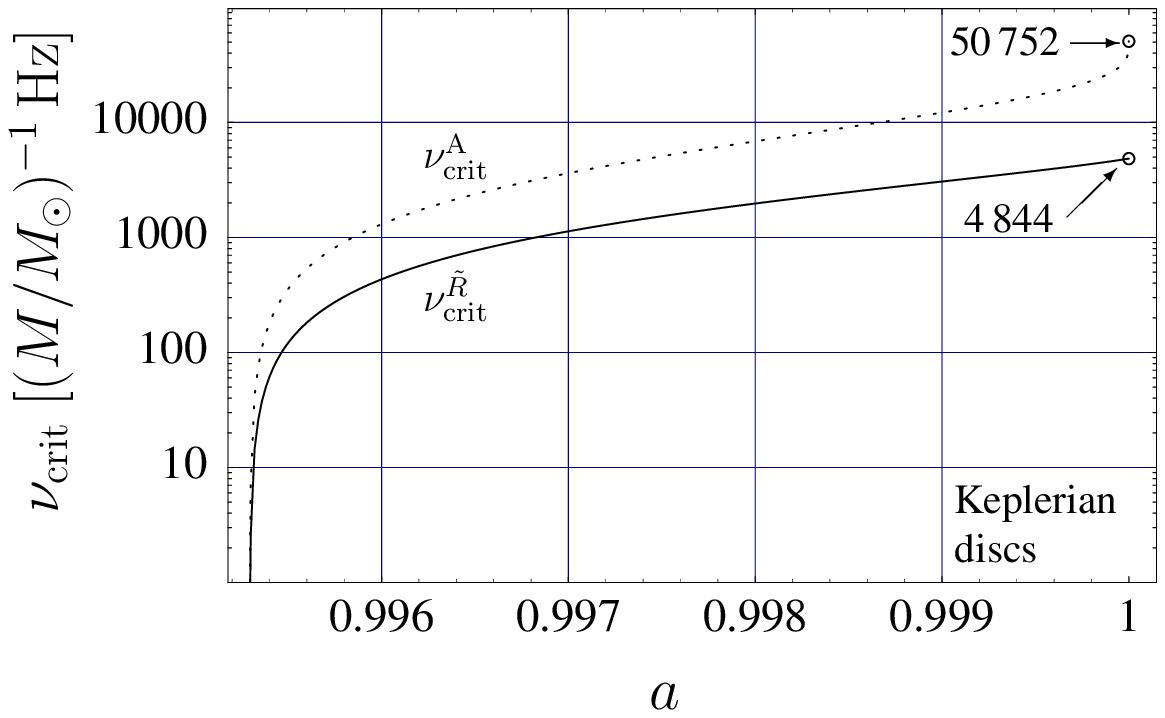}
\end{minipage}
\begin{minipage}{.49 \hsize}
\centering
\includegraphics[width=.95 \hsize]{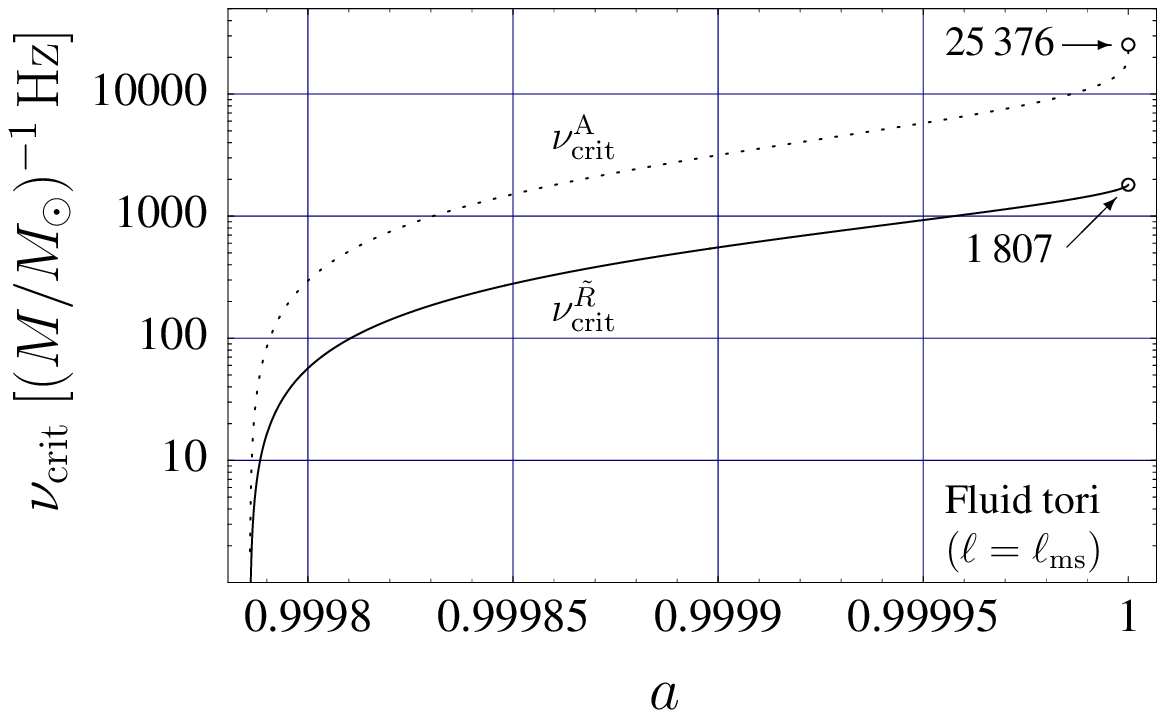}
\end{minipage}
\caption{Critical frequency $\nu^{A}_{\rm crit}$
  defined in terms of the B--L coordinate radius \citep{Asch:2004:ASTRA:} and
  the physically correct (coordinate independent) critical frequency
  $\nu^{\tilde{R}}_{\rm crit}$ defined in terms of the proper radial distance,
  as a function of the rotational parameter $a$ of the black hole.}
\label{f5}
\end{figure*}

\begin{figure*}
  \centering
  \includegraphics[width=.78 \hsize]{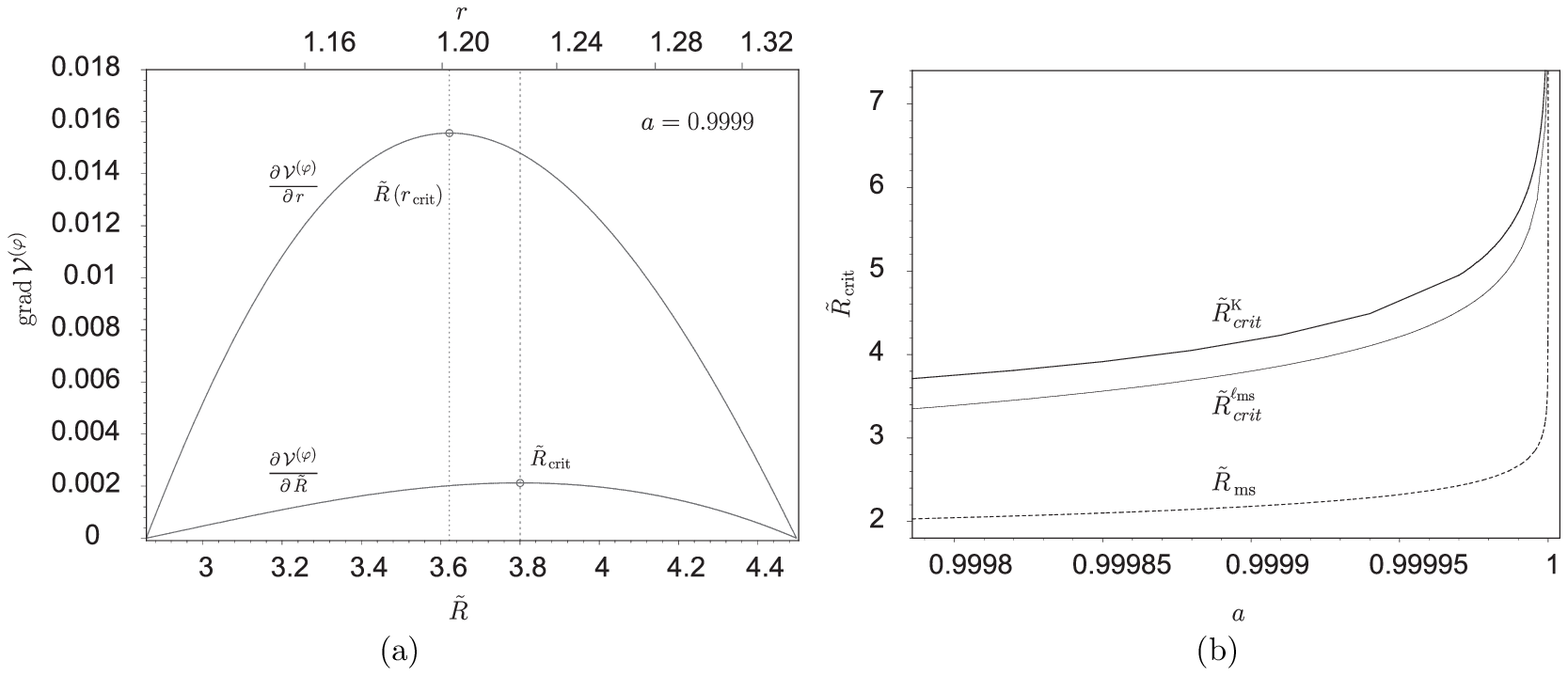}
  \caption{Determination of the critical ``humpy'' frequency.
  (a) Positive parts of the ``coordinate'' and ``proper'' radial
  gradient $\p {\cal V}^{(\phi)}/\p r$ and $\p {\cal V}^{(\phi)}/\p
  \tilde{R}$ for a given value of the rotational parameter $a$ in the
  Keplerian disc.
  (b) Proper radial distance of the loci of $(\p {\cal V}^{(\phi)}/\p
  \tilde{R})_{\rm max}$ measured from the marginally bound orbit for both the
  Keplerian discs ($\tilde{R}_{\rm crit}^{\rm K}$) and $\ell=\ell_{\rm ms}$
  perfect-fluid tori ($\tilde{R}_{\rm crit}^{\ell_{\rm ms}}$). Proper radial
  distance to the marginally stable orbit ($\tilde{R}_{\rm ms}$) is also
  shown.} 
\label{f6}
\end{figure*}

In terms of the redefined rotational parameter $(1-a)$, the ``humpy'' profile
of the LNRF orbital velocity of marginally stable thick discs occurs
for discs orbiting Kerr black holes with $(1-a)<1-a_{\rm c(T)}\doteq
2.1\times 10^{-4}$, which is more than one order lower than
the value $1-a_{\rm c(K)}\doteq 4.7\times 10^{-3}$ found by
\citet{Asch:2004:ASTRA:} for the Keplerian thin discs. Moreover, in the thick
discs, the velocity difference $\Delta {\cal V}^{(\phi)}_{\mathrm{T}}$
is smaller but comparable with those in the thin discs (see Fig.~\ref{f3}). In
fact, we can see that for $a \to 1$, the velocity difference in
the thick discs $\Delta {\cal V}^{(\phi)}_{\rm (T)}\approx 0.02$, while 
for the Keplerian discs it goes even up to $\Delta {\cal
  V}^{(\phi)}_{\rm(K)}\approx 0.07$. 

\section{Humpy frequency and its relation to epicyclic frequencies}
\label{section3} 

In Kerr spacetimes, the frequencies of the radial and latitudinal
(vertical) epicyclic oscillations related to an equatorial Keplerian circular
orbit at a given $r$ are determined by the formulae
\citep[e.g.,][]{Ali-Gal:1981:GENRG2:,Now-Leh:1998:TheoryBlackHoleAccretionDisks:}
\bea
        \nu^2_{\rm r} &=& \nu^2_{\rm K}(1-6r^{-1} + 8ar^{-3/2} - 3a^2r^{-2}),
        \quad \\ 
        \nu^2_{\rm v} & \equiv & \nu^2_{\theta} = \nu^2_{\rm K} (1-
        4ar^{-3/2} + 3a^2r^{-2}), 
\eea
where $\nu_{\rm K} = \Omega_{\rm K}/2\pi$. A detailed analysis of properties
of the epicyclic frequencies can be found in
\citet{TOR-STU:2005:RAG,Tor-Stu:2005:Radial}. The epicyclic oscillations with
the frequencies $\nu_{\rm r},\ \nu_\mathrm{v}$ can
be related to both the thin Keplerian discs \citep{Abr-Klu:2000:ASTRAL:,
  Kat:2004:PUBASJ:mass} and thick, toroidal discs \citep{Rez-etal:2003:MONNR:,
  Klu-Abr-Lee:2004:X-RayTiming2003:}. 

\cite{Asch:2004:ASTRA:,Asch:2006:ASTROPH:} defined 
the characteristic (critical) frequency of any related mechanism
possibly exciting the 
disc oscillations in the region of positive gradient of its
LNRF-velocity ${\cal V}^{(\phi)}$ by the maximum positive slope of ${\cal
  V}^{(\phi)}$: 
\be                                                        \label{eq22}
     \nu^{A}_{\rm crit}=\left.\frac{\p {\cal V}^{(\phi)}_\mathrm{K}}{\p
         r}\right|_{\rm max}.
\ee
This frequency has to be determined numerically and we have done
it for both the Keplerian discs and the marginally stable discs with
$\ell=\ell_{\rm ms}=\mathrm{const}$, see Fig.~\ref{f5} and Table \ref{t1}. 

Although there is no detailed idea on the mechanism generating the
``hump-induced'' oscillations, it is clear that the Aschenbach proposal of
defining the characteristic frequency deserves attention. 
It should be stressed, however, that a~detailed analysis of the instability
could reveal a~difference between the characteristic frequency and the actual
observable one, as the latter should be associated with the fastest growing
unstable mode\footnote{We thank to the referee for pointing out this
  possibility.}.
Moreover the frequency $\nu_{\rm crit}^{A}$, defined by Eq. (\ref{eq22}),
represents an upper limit on the frequencies of the hump-induced oscillations,
as it is given by maximum of the LNRF-velocity gradient in the humpy part of
the velocity profile. 

In the following we assume that the characteristic (critical) frequency is a
typical frequency of oscillations induced by the conjectured ``humpy
instability'', and that the humpy oscillations could excite oscillations with
the epicyclic frequencies or some combinational frequencies, if appropriate
conditions for a~forced resonance are satisfied in vicinity of the radius where
the humpy oscillations occur.

\begin{figure}
  \centering
  \includegraphics[width=1 \hsize]{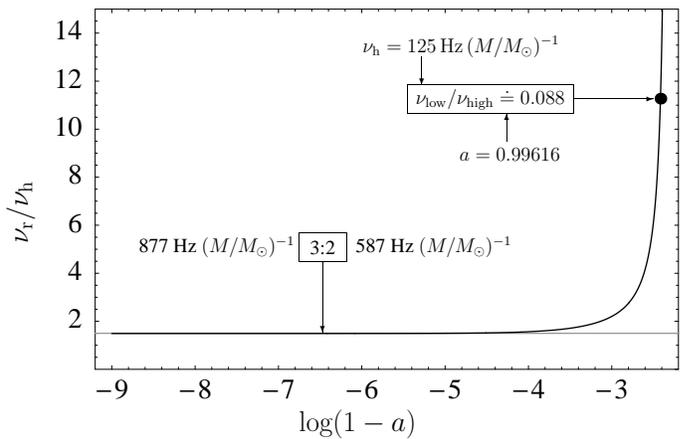}
  \caption{Spin dependence of the ratio of the radial epicyclic frequency and
  the ``humpy frequency'' related to distant observers. The ratio is given in
  the radius of definition of the humpy frequency $r_{\rm h}$. In the
  interval of $1-a \in 
  (1.7\times 10^{-3},10^{-4})$, the ratio rapidly falls down, to the asymptotic
  value of $3:2$ starting at $a \sim 10^{-4}$. Then an exact $1/\mathrm{M}$
  scaling holds with frequencies depicted in the figure. Notice that at the
  Aschenbach's value of $a\simeq 0.99616$, for which the resonant orbit with
  $\nu_{\rm v}:\nu_{\rm r}\sim 3:1$ is close to $r_{\rm h}$, there is
  $\nu_{\rm r}/\nu_{\rm h} \sim 12$, analogous to the ratio of high and low
  frequency QPOs.}
\label{f7}
\end{figure}

In situations where the general relativity is crucial, it is necessary
to consider $\p {\cal V}^{(\phi)}/\p \tilde{R}$, where $\tilde{R}$ is the
physically relevant (coordinate-independent) proper radial distance, as this
is an appropriate way for estimating the characteristic
frequencies related to local physics in the disc. Then correct
general relativistic definition of the critical frequency for possible
excitation of oscillations in the disc is given by the relations
\be                                                        \label{eq23}
     \nu^{\tilde{R}}_{\rm crit} = \left.\frac{\p {\cal V}^{(\phi)}}{\p
       \tilde{R}}\right|_{\rm max}, \quad
     \d \tilde{R} = \sqrt{g_{rr}}\d r = \sqrt{\frac{\Sigma}{\Delta}}\d
     r\ensuremath{\mathrm{,}}
\ee
where ${\cal V}^{(\phi)} = {\cal V}^{(\phi)}_\mathrm{K}(r;\,a)$ in thin
Keplerian discs, and ${\cal V}^{(\phi)} = {\cal
  V}^{(\phi)}_\mathrm{T}(r;\,l,\,a)$ in marginally stable thick discs. 
Of course, such a locally defined frequency, confined naturally to the
observers orbiting the black hole with the LNRF, should be further related to
distant stationary observers by the formula (taken at the B--L
coordinate $r$ corresponding to $(\p {\cal V}^{(\phi)}/\p \tilde{R})_{\rm
  max})$ 
\be                                                           \label{rce_24}
       \nu_{\rm h}=\nu^{\tilde{R}}_{\infty} = \sqrt{-(g_{tt}+2\omega
       g_{t\phi}+\omega^2 g_{\phi\phi})} \,\nu^{\tilde{R}}_{\rm crit}.
\ee
We suggest to call such a coordinate-independent and, in principle,
  observable frequency the ``humpy frequency'', as it is related to the humpy
  profile of ${\cal V}^{(\phi)}$, and denote it $\nu_{\rm h}$.
Again, the physically relevant humpy frequency $\nu_{\rm h} =
\nu^{\tilde{R}}_{\infty}$, connected to observations by distant observers and
exactly defined by Eqs. (\ref{eq23}) and (\ref{rce_24}), represents an upper
limit on characteristic frequencies of oscillations induced by the hump of the
LNRF-velocity profile, and the realistic humpy frequencies, as observed
by distant observers, can be expected close to but smaller than
$\nu^{\tilde{R}}_{\infty}$.
Further, we denote $r_{\rm h}$ the B-L radius of definition of the humpy
oscillations frequency, where
$\p {\cal V}^{(\phi)}/\p \tilde{R} = (\p {\cal V}^{(\phi)}/\p \tilde{R})_{\rm
  max}$. Of course, in realistic situations the hump-induced oscillation
mechanism could work at the vicinity of $r_{\rm h}$, with slightly different
frequencies; we should take into account that the shift of the radius, where
the mechanism works, shifts both the locally measured (LNRF) frequency
(Eq.~(\ref{eq23})) and the frequency related to distant observers
(Eq.~(\ref{rce_24})). The zones of radii, where the critical frequency
$\nu^{\tilde{R}}_{\rm crit}$ differs up to 1\%, 10\% and 20\% of its maximal
value (given by $(\p {\cal V}^{(\phi)}/\p \tilde{R})_{\rm max}$) for thin
(Keplerian) discs or 1\%, 5\% and 10\% of its maximum for marginally stable
discs with $\ell=\ell_{\rm ms}$, are given in Fig.~\ref{f4}.

\begin{figure}
  \centering
  \includegraphics[width=.95 \hsize]{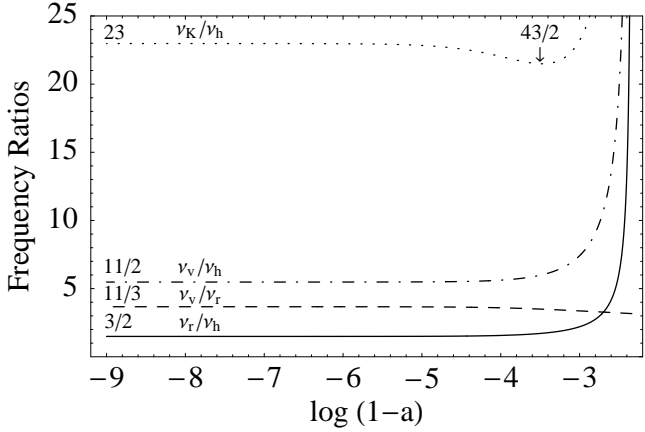}
  \caption{Spin dependence of the ratios of the radial ($\nu_{\rm r}$) and
  vertical ($\nu_{\rm v}$) epicyclic frequencies, and the Keplerian frequency
   ($\nu_{\rm K}$) to the thin-disc humpy frequency related to distant
   observers ($\nu_{\rm h}$). Further the ratio of the epicyclic frequencies 
   is given at the radius of definition of the humpy frequency. All the
   frequency ratios are asymptotically (for $1-a<10^{-4}$) constant. There si
   $\nu_{\rm K}:\nu_{\rm v}:\nu_{\rm r}:\nu_{\rm h}\sim 46:11:3:2$. Therefore,
   we can expect some resonant phenomena on the ratio of 
   $\nu_{\rm r}:\nu_{\rm h}\sim 3:2$, and $\nu_{\rm K}:\nu_{\rm v}\sim 4$
   that could be both correlated.} 
\label{f8}
\end{figure}

An analogical relation to Eq. (\ref{rce_24}) can be written also for the
Aschenbach critical frequency  $\nu^{\rm A}_{\rm crit}$, giving the Aschenbach
frequency related to distant observers $\nu^{\rm A}_{\infty}$. Because the
velocity gradient related to the proper distance $\tilde{R}$ is suppressed in
comparison with that related to the Boyer-Lindquist coordinate distance $r$,
there is $\nu^{\tilde{R}}_{\rm crit} < \nu^{\rm A}_{\rm crit}$. The situation
is illustrated in Fig.~\ref{f5}. Moreover, Fig.~\ref{f6} shows mutual
behaviour of the ``coordinate'' and ``proper'' radial gradient $\p {\cal
  V}^{(\phi)}/\p r$ and $\p {\cal V}^{(\phi)}/\p \tilde{R}$ in the region
between the local minimum and the outer local maximum of the orbital velocity
${\cal V}^{(\phi)}$ of $\ell=\ell_{\rm ms}=\mbox{const}$ discs for an
appropriately chosen value of the rotational parameter $a$. It is interesting
to compare the Aschenbach frequencies (defined in terms of the B-L coordinate
$r$) with the critical frequencies defined in terms of the proper radial
distance $\tilde{R}$. Characteristic frequencies $\nu^{\rm A}_{\rm crit},\
\nu^{\rm A}_{\infty},\ \nu^{\tilde{R}}_{\rm crit},\ \nu^{\tilde{R}}_{\infty}$
are given in Table~\ref{t1} for some typical values of the rotational
parameter $a$ for both Keplerian discs and limiting $\ell=\mbox{const}$ tori
with $\ell=\ell_{\rm ms}$.

\begin{figure}
  \centering
  \includegraphics[width=1 \hsize]{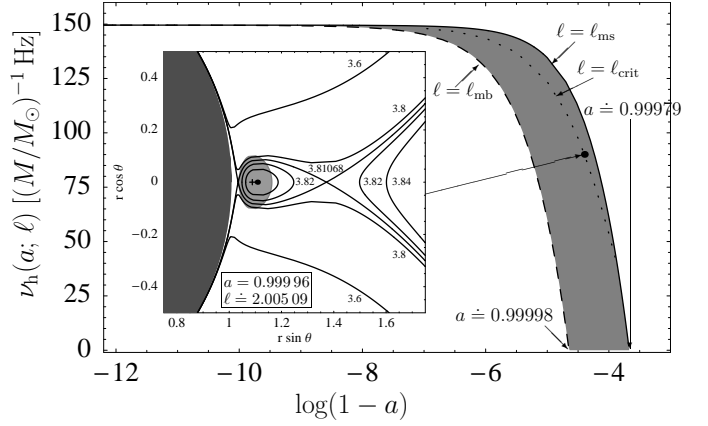}
  \caption{Interval of humpy frequencies for the marginally stable
  thick discs with $\ell\in (\ell_{\rm ms},\,\ell_{\rm
  mb})$ as a function of the \bh spin $a$. For $a\to 1$, the interval is
  narrowing and asymptotically reaching the value of $150\,
  \mbox{Hz}\,(M/M_{\odot})^{-1}$. Dotted curve corresponds to the humpy 
  frequencies of marginally stable slender tori with
  $\ell=\ell_{\rm crit}$, for which the critical von~Zeipel surface contains
  two cusps (as it is demonstrated for one special case in the left panel of
  the figure; the thick part of the torus is given by the light-gray region).}
\label{f9}
\end{figure}

\begin{figure}
  \centering
  \includegraphics[width=1 \hsize]{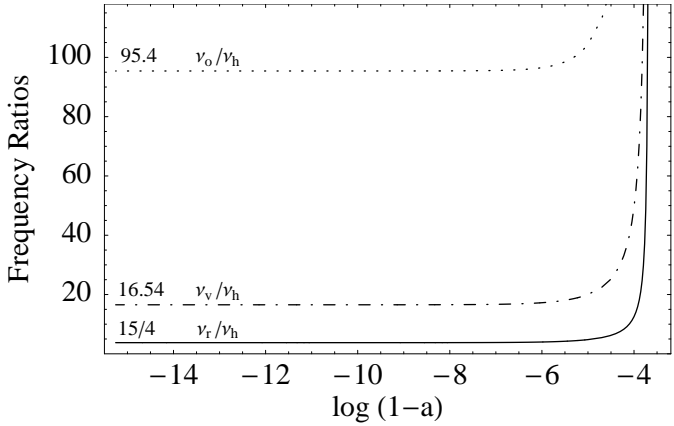}
  \caption{Spin dependence of the ratios of the radial ($\nu_{\rm r}$) and
  vertical ($\nu_{\rm v}$) epicyclic frequencies, and the orbital frequency
   ($\nu_{\rm o}$) of the marginally stable $\ell=\ell_{\rm ms}$ disc to the
   thick-disc humpy frequency related to distant observers ($\nu_{\rm h}$).
  All the frequency ratios are asymptotically (for $1-a<10^{-6}$) almost
   constant.} 
\label{f10}
\end{figure}

The physically and observationally relevant frequency connected to the
LNRF-velocity gradient sign change is given by the frequency $\nu_{\rm
  h}=\nu_{\infty}^{\tilde{R}}$ corresponding to the locally ``hump-induced''
oscillations taken from the point of view of distant stationary observers. In
order to obtain an intuitive insight into a possible observational relevance
of $\nu_{\rm h}$, it is useful to compare it with the 
frequencies of the radial and vertical epicyclic oscillations, $\nu_{\rm
  r}$ and $\nu_{\rm v}$, and the orbital frequency of the disc, $\nu_{\rm orb}
  =\Omega/2\pi$, where $\Omega$ is given for both thin and thick discs by
  Eq.~\ref{e4} and the appropriate distribution of the specific angular
  momentum $\ell$. The most interesting and crucial phenomenon is the spin
  independence of the frequency ratios for extremely rapid Kerr black holes.
The results are given in Figs.~\ref{f7}--\ref{f10}.
Further we can see
(Figs.~\ref{f4}) that the resonant epicyclic frequencies radii $r_{3:1}$
and $r_{4:1}$ are located within the zone of the hump-induced
oscillation mechanism in both thin discs and marginally stable tori.

We would like to call attention to the fact that in Keplerian discs the sign
changes of the  
radial gradient of the orbital velocity in LNRF occur nearby the $r=r_{3:1}$
orbit (with $\nu_{\rm v}:\nu_{\rm r}=3:1$), while in the vicinity of the
$r=r_{3:2}$ orbit (with $\nu_{\rm v}:\nu_{\rm r}=3:2$), $\p {\cal
  V}^{(\phi)}/\p
r<0$ for all values of $a$ for both Keplerian discs and marginally
stable tori with all allowed values of $\ell$. 
The parametric resonance, which is the strongest one for the ratio of the 
epicyclic frequencies $\nu_{\rm v}:\nu_{\rm r}=3:2$, can occur at the
$r=r_{3:2}$ orbit, while its effect is much smaller at 
the radius $r=r_{3:1}$, as noticed by \cite{Abr-etal:2003:PUBASJ:}. 
Nevertheless, the forced resonance may take place at the $r_{3:1}$
  orbit. 
Notice that the forced resonance at $r=r_{3:1}$ can generally result in
observed QPOs 
frequencies with 3:2 ratio due to the beat frequencies allowed for the forced
resonance as shown in \cite{Abr-etal:2004:RAGtime4and5:}. 
But the forced resonance at $r_{3:1}$ between the epicyclic frequencies,
induced by the humpy profile of ${\cal V}^{(\phi)}$, seems to be irrelevant in
the case of microquasars, since all observed frequencies lead 
to the values of the rotational parameter $a<a_{\rm c(K)}$, as shown by
\cite{Tor-Abr-Klu-Stu:2005:ASTRA:}. On the other hand, the 
LNRF-velocity hump could induce the forced resonance between another
(non-epicyclic) frequencies as well, and thus being relevant also for
microquasars like the nearly extreme Kerr black hole candidate GRS~1915$+$105
\citep{McCli-etal:2006:ASTROPH:}.

The marginally stable tori have a structure that depends on the value of the
specific angular momentum $\ell\in(\ell_{\rm ms},\,\ell_{\rm mb})$. The
oscillations of slender tori ($\ell\approx\ell_{\rm ms}$) have frequencies
equal to the epicyclic frequencies relevant for test particle motion, but the
frequencies of non-slender tori are different, as shown for pseudo-Newtonian
tori \citep{Sra:2005:ASTRN:,Bla-etal:2006:ASTRJ2:} and expected for tori in
the strong gravitational field of Kerr black holes. Therefore, comparison of
the humpy frequencies and the epicyclic frequencies is relevant for the
slender tori only.

The humpy frequency is defined for all $a>0.99979$ and all $\ell\in(\ell_{\rm
ms},\,\ell_{\rm mb})$, see Fig.~\ref{f9}. It is important that in the field
of Kerr black holes with $1-a<10^{-8}$, there is $\nu_{\rm
  h}(a,\,\ell)\simeq 150\, 
\mbox{Hz}\,(M/M_{\odot})^{-1}$ independently of $a$ and $\ell$. Further, it
is shown that physically important case of tori admitting evolution of
toroidal von~Zeipel surfaces with the critical surface self-crossing in both
the inner and the outer cusps is allowed at $\ell=\ell_{\rm crit}$, where
$\ell_{\rm crit}\gtrsim\ell_{\rm ms}$ only slightly differs from $\ell_{\rm
ms}$, i.e., such tori can be slender, see Fig.~\ref{f9}. The ratios of
$\nu_{\rm r}/\nu_{\rm h},\ \nu_{\rm v}/\nu_{\rm h}$ and $\nu_{\rm o}/\nu_{\rm
  h}$ are given for the tori with $\ell\approx\ell_{\rm ms}$ in
Fig.~\ref{f10}. Their asymptotical values, valid for $1-a<10^{-6}$, are
independent of both $a$ and $\ell$.

\section{Concluding remarks}\label{section4}

\begin{table*}
\centering
\caption{Characteristic frequencies in units of
  $(M/M_{\odot})^{-1}$\,Hz ($M/M_{\odot}$ is the mass of the Kerr black hole
  in units of mass of the Sun), corresponding to critical frequencies
  $\nu^{\rm A}_\mathrm{crit}$, $\nu_{\infty}^{\mathrm{A}}$,
  $\nu^{\tilde{R}}_\mathrm{crit}$, $\nu^{\tilde{R}}_{\infty}$  defined in the
  text, are given for appropriate values of the 
  black-hole spin. Maximal values of the frequencies related to the stationary
  observer at infinity are bold-faced. 
Note that only the frequencies $\nu_\infty^{\tilde{R}}$ have physical
  meaning for direct comparison with the frequencies of orbital oscillations
  $\nu_{\rm orb},\,\nu_{\rm r},\,\nu_{\rm v}$ related to the
  observers at infinity.
} 
\begin{tabular}{rcccccccccc}
\hline\hline
\rule{0pt}{10pt}
& & \multicolumn{4}{c}{Keplerian discs} & & \multicolumn{4}{c}{Fluid tori} \\
\multicolumn{1}{c}{$1-a$} & & $\nu^{\rm A}_\mathrm{crit}$ & $\nu^{\rm
  A}_{\infty}$ & $\nu^{\rm \tilde{R}}_\mathrm{crit}$ & 
$\nu^{\rm \tilde{R}}_{\infty}$ & & $\nu^{\rm A}_\mathrm{crit}$ & $\nu^{\rm
  A}_{\infty}$ & $\nu^{\rm\tilde{R}}_\mathrm{crit}$ & 
$\nu^{\rm \tilde{R}}_{\infty}$ \\[.5ex]
\hline
\rule{0pt}{10pt}
$4.5\times 10^{-3}$ & & 356 & 86 & 121 & 29 & & & & \\
$4\times 10^{-3}$ & & 1303 & 303 & 432 & 102 & & & & \\
$3\times 10^{-3}$ & & 3617 & 767 & 1130 & 248 & & \multicolumn{4}{c}{\em
  undefined} \\
$1\times 10^{-3}$ & & 12179 & 1849 & 3061 & 536 & & & & \\
$5\times 10^{-4}$ & & 17132 & 2126 & 3789 & 592 & & & & \\
$2\times 10^{-4}$ & & 22982 & {\bf 2203} & 4352 & {\bf 607} & & 296 & 34 & 57
& 7 \\ 
$1\times 10^{-4}$ & & 26857 & 2126 & 4579 & 603 & & 3160 & 315 & 555 & 61 \\
$1\times 10^{-5}$ & & 36593 & 1565 & 4816 & 590 & & 10940 & {\bf 657} & 1447 &
135 \\ 
$1\times 10^{-6}$ & & 42556 & 1001 & 4841 & 588 & & 16271 & 589 & 1718 & 147 \\
$1\times 10^{-9}$ & & 49250 & 201 & 4844 & 588 & & 23277 & 185 & 1807 & {\bf
  150} \\ 
\hline
\end{tabular}
\label{t1}
\end{table*}

The equality of $\nu^{\rm A}_{\rm crit}$ and $\nu_{\rm r}$ for the Kerr black
holes with $a \simeq 0.99616$, indicating direct relation of the Aschenbach
characteristic frequency and the radial epicyclic frequency
\citep{Asch:2004:ASTRA:,Asch:2006:ASTROPH:}, is rather
only an accidental coincidence, because $\nu^{\rm A}_{\rm crit}$ is defined in
a~coordinate-dependent way. The physically relevant frequency
$\nu_\mathrm{h}=\nu^{\tilde{R}}_{\infty}$ cannot be directly related to the
radial epicyclic frequency in Keplerian discs, as
$\nu^{\tilde{R}}_{\infty}<\nu_{\rm r}$ for all relevant values of $a \in
(a_{\rm c(K)},\,1)$.
Nevertheless, the behaviour of the ratio $\nu_{\rm
  r}/\nu_{\infty}^{\tilde{R}}$ indicates some interesting consequences (see
Fig.~\ref{f7}). 

First, for thin (Keplerian) discs around the Kerr black holes with $a
\simeq 0.99616$, when 
the ratio of epicyclic frequencies $\nu_{\rm v}:\nu_{\rm r} \sim 3:1$ at
the radius of definition of $\nu_{\infty}^{\tilde{R}}$, we find $\nu_{\rm
  r}:\nu_{\infty}^{\tilde{R}} \sim (11$--$13):1$, i.e., in such a~situation 
the frequency induced by the positive gradient of the LNRF-velocity profile
could be related to the low-frequency oscillations. 
However, such explanation is restricted to 
extremely rapidly rotating black holes and, contrary to the idea of the
13-th wave \citep{Abr-etal:2004:RAGtime4and5:}, cannot be extended to other
black-hole, neutron-star, and white-dwarf systems.
Therefore, this has to be taken as a kind of curiosity working for a very
special class of black-hole systems only. 

Second, for thin (Keplerian) discs around the Kerr black holes
with $a>0.9999$, there is the
ratio of $\nu_{\rm r}:\nu_{\infty}^{\tilde{R}} \sim 3:2$, and
$\nu_{\rm v}:\nu_\infty^{\tilde{R}} \sim 11:2$, independently of
$a$. Assuming that the oscillations at the humpy frequency $\nu_\mathrm{h} =
\nu^{\tilde{R}}_\infty$ could be really directly detected by distant
observers, for such black holes with $1-a<10^{-4}$ the high-frequency twin
peak QPOs with $3:2$ ratio could be explained independently of the standard
resonant phenomena, if we focus on the asymptotic behaviour of
$\nu_\mathrm{r}:\nu_\infty^{\tilde{R}}\sim 3:2$. Moreover, for such extremely
rapid Kerr black holes with $1-a<10^{-4}$, we could consider triples of
frequencies
taken in rational ratios $\nu_{\rm v}:\nu_{\rm r}:\nu_\infty^{\tilde{R}}\sim
11:3:2$, if the epicyclic oscillations are excited by the LNRF-velocity hump.
Such frequency ratios could be observed mainly in disc systems
around supermassive black holes in galactic nuclei that are expected to be 
extremely fast rotating; especially Sgr A* should be tested very carefully for
this possibility. For Kerr black holes with the spin parameter
$1-a>10^{-4}$, the frequency ratio is different and depends strongly
on the spin $a$ (see Figs. \ref{f7} and \ref{f8}). 

Considering also the Keplerian frequency we find the ratio of
$\nu_{\mathrm{K}}:\nu_{\infty}^{\tilde{R}}$ having a local minimum for $a
\simeq 0.99965$ and a nearly constant value 
for $1-a<10^{-5}$, where
$\nu_\mathrm{K}:\nu_\infty^{\tilde{R}}\sim 23$. In the field of Kerr black
holes with $1-a<10^{-5}$, the frequency ratios
$\nu_\mathrm{K}:\nu_{\rm v}:\nu_{\rm r}:\nu_\infty^{\tilde{R}} \sim
46:11:3:2$ are almost independent of $a$.
Thus for the extremely rapid Kerr black holes the $1/M$ scaling of considered
frequencies is quite exact.
Note that in such a case there is
$\nu_\mathrm{r}:\nu_\mathrm{h}\sim 3:2$ 
and the ratio $\nu_{\rm K}:\nu_{\rm v}$ is close to the ratio 4:1
at the radius of definition of the humpy frequency. This indicates a
possibility of ``doubled'' resonant phenomena with the special frequency
  ratios in Keplerian discs orbiting extremely rapid Kerr black holes
  ($1-a<10^{-5}$).

Third, the hump-induced oscillations with frequencies 
$\nu_{\rm h}\lesssim\nu^{\tilde{R}}_{\infty}$
could be generated in a zone around $r_{\rm h}$ ($\nu_{\rm
  h}=\nu^{\tilde{R}}_{\infty}$ at $r_{\rm h}$), where the resonant phenomena
between 
the radial and vertical epicyclic oscillations could enter the game, namely at
the ratios of $\nu_{\rm v}:\nu_{\rm r}=3:1\ \mbox{and}\ 4:1$. Interesting
resonant phenomena could be then expected when the $\nu_{\rm r}:\nu_{\rm h}$ 
corresponds to the ratio of small integer numbers. Especially the case of
$\nu_{\rm r}:\nu_{\rm h}\sim 3:2$ in spacetimes with $1-a<10^{-4}$ is worth of
attention.
In general, observationally relevant should be the
resonances represented by frequency ratios in small integer numbers $p:q$. As
shown in \citet{Lan-Lif:1973:Mech:}, the relevance of resonant phenomena
depends on the order of resonance $n=\mbox{max}(p,\,q)$, and falls steeply (in
powers) with increasing value of $n$; in fact they argue that relevant
resonant phenomena could be expected for $n\leq 4$.
Therefore, the frequency ratios such as 23:1, 11:2, 11:3 appear to be quite
irrelevant in realistic resonance models.

Recall that there is a~well known Thorne limit giving the maximum spin of the
Kerr black hole in systems with thin accretion discs, $a_{\rm max}\simeq
  0.998$, determined by the back-reaction of photons radiated from the disc
  and captured by the black hole
  \citep{Pag-Tho:1974:ASTRJ2:,Tho:1974:ASTRJ2:}. If the hump-induced
  oscillations and related epicyclic frequencies will be observed in ratios 
corresponding to the asymptotic region of $a>0.9999$ for Keplerian discs, the
Thorne model should be corrected, e.g., by effect of an occultation of the
disc. In the case for which the Thorne limit turns out to be realistic, the
hump-induced oscillations have to be restricted on the spin interval
$a\in (0.9953,\,0.998)$. We expect the Thorne limit being relevant for smooth
thin discs, while the overcoming of $a_{\rm max}$ could be expected in
highly turbulent discs with toroidal internal parts. 

For thick discs the situation is much more complex, being dependent on
both the rotational parameter (spin) $a$ and the specific angular
momentum $\ell$. 
The range of maximal humpy frequencies for a given spin $a$ is plotted in
Fig. \ref{f9} 
and is determined by their evaluation in limiting values of the specific
angular momentum $\ell$ relevant for the ``humpy'' effect in marginally stable
thick accretion discs (see the discussion in Sec. \ref{s2.2}). The minimal
value corresponds to $\ell_{\rm ms}(a)$ while the maximal value, in dependence
of $a$, corresponds to $\ell_{\rm ex(max)}(a)$ (for $0.99979\leq a \leq
0.99998$) or $\ell_{\rm mb}(a)$ (for $0.99998\leq a \leq 1$). Notice that
asymptotically (for $1-a<10^{-8}$) both $\nu_{\rm h(ms)}$ and $\nu_{\rm
  h(mb)}$ coincide on the line of 150 Hz $(M/M_{\odot})^{-1}$. Clearly, the
same is true for the humpy frequencies related  
to discs with any relevant $\ell\in(\ell_{\rm ms},\, \ell_{\rm mb})$. The
spin-dependence of the ratio of the humpy frequency and the epicyclic and
orbital frequencies (taken at the radius of definition of the humpy frequency)
for the case of limiting $\ell=\ell_{\rm ms}$ discs is given in
Fig.~\ref{f10}. Again we obtain asymptotically constant (spin-independent)
ratios for black holes with $1-a<10^{-6}$, where $\nu_{\rm r}:\nu_{\rm
  h(ms)}\sim 15:4,\ \nu_{\rm v}:\nu_{\rm r}\doteq 4.39,\ \nu_{\rm v}:\nu_{\rm
  h(ms)}\doteq 16.54,\ \nu_{\rm orb}:\nu_{\rm r}\doteq 25.4$, and
  $\nu_{\rm orb}:\nu_{\rm h}\doteq 95.4$. It should be  
stressed that for the holes with $1-a<10^{-6}$ the same ratios with the humpy
frequencies are obtained for the discs with any $\ell\in(\ell_{\rm ms},\,
\ell_{\rm mb})$, as $\ell_{\rm mb}\to\ell_{\rm ms}\to 2$ for $a\to 1$. 
The asymptotically constant values of the frequency ratios correspond to the
rational value only in the case 
of $\nu_{\rm r}:\nu_{\rm h}\sim 15:4$. Of course, we could find some rational
ratios for any ones, if $1-a>10^{-6}$. On the other hand, we directly see
(Fig.~\ref{f4}b) that for the very slender marginally stable tori
($\ell\approx\ell_{\rm ms}$) the resonant phenomena on epicyclic frequencies
with $\nu_{\rm v}:\nu_{\rm r}=4:1$ ratio appear in very close vicinity of the
humpy radius $r_{\rm h}$, making thus a very special prediction on the QPOs
frequencies observed in such hypothetical systems with Kerr black holes
having $1-a<2\times 10^{-4}$. 
In the marginally stable slender tori the resonant phenomena between the
radial epicyclic and humpy oscillations, taking place at the humpy radius
$r_{\rm h}$, and between the vertical and radial epicyclic oscillations near
the humpy radius, both with the ratio $\sim 4:1$, could be observationally
relevant only, but their relevance is expected to be lower than that of
the frequency ratios 3:2 and 3:1 in Keplerian discs.

Finally, it should be stressed that at present no direct mechanism triggering
the LNRF velocity hump excited oscillations is known, being a challenge for
investigation, since the existence of the toroidal von~Zeipel surfaces (see
Figs.~\ref{f2},~\ref{f9}) brings some indication of possible triggering of 
instabilities in both radial and vertical directions leading to oscillations
in accretion discs. The predictions for the ratio of the humpy and epicyclic
or Keplerian (orbital) frequencies presented here for both thin discs
and slender tori have to be compared with observations made in nearly extreme
Kerr black hole systems. In the case of the humpy oscillations excited systems
we could observe more than two QPOs with frequencies in rational ratio. 
It seems that in the X-ray variable binary system (microquasar) with the
nearly extreme Kerr black hole candidate GRS~1915$+$105 four oscillations with
related frequencies have been observed, what brings a large field for testing
the predictions of the ``LNRF-velocity hump excited oscillations''
model. The tests have to be done in a close connection to both the related
resonance model and the results of the spectral analysis of the X-ray
continuum, as observed in GRS~1915$+$105
\citep{Rem:2005:Xray:,McCli-etal:2006:ASTROPH:}
\footnote{
If the results of \citet{Asch-etal:2004:X-ray} will be confirmed by more
precise observations, the Galactic Center black hole system Sgr~A* could serve
as another example of the nearly extreme Kerr black hole system with more than
two QPO oscillations observed that could test the ``humpy'' model.}.
We believe that a synergy effect of such studies could lead
to deeper understanding of X-ray binary systems, namely microquasars.

\begin{acknowledgements}

This work was supported by the Czech grant MSM~4781305903.
\end{acknowledgements}

\bibliographystyle{aa} 


\end{document}